\documentclass[aps,pra,twocolumn,superscriptaddress,showpacs]{revtex4-1}

\usepackage{amssymb}
\usepackage{color,graphicx}
\usepackage{amsmath}
\usepackage{amsbsy}
\usepackage{amsthm}
\usepackage{bbm}
\usepackage{bm}
\usepackage{epsfig}
\usepackage{palatino}
\usepackage{float}
\usepackage{subfigure}

\usepackage[]{microtype}
\emergencystretch=\maxdimen
\usepackage{graphicx}
\usepackage{dcolumn}
\usepackage{bm}
\usepackage{braket,palatino}
\usepackage{bbm}

\begin{document}

\title{Discrete-time quantum walks as generators of multipartite entanglement}%
\author{J. Lockhart}%
\thanks{JL's current address: Department of Computer Science, University College London, London, WC1E 6BT, UK}
\affiliation{%
	School of Electronics, Electrical Engineering and Computer Science, \\Queen's University, Belfast, BT7 1NN, UK
}%
\affiliation{ %
	Department of Physics and Astronomy, University College London, London, WC1E 6BT, UK
} %
\author{M. Paternostro}%
\affiliation{%
 Centre for Theoretical Atomic, Molecular and Optical Physics, School of Mathematics and Physics, Queen's University, Belfast, BT7 1NN, UK
}%

\date{\today}

\begin{abstract}
We extend the idea of a discrete-time quantum walk on a graph by placing a qubit on each vertex, and allowing the walker to interact with the qubit at its current position. We show that allowing for a controlled-Z interaction at each time step between the walker and the vertex qubits generates multipartite entanglement between the qubits. We demonstrate that for particular coin operators the system generates a highly entangled cluster state, of use in measurement based quantum computation.
\end{abstract}
\maketitle
\section{\label{sec:introduction}Introduction}
Quantum walks~\cite{Kempe}, the quantum analogue of classical random walks, are conceptually simple protocols that display a rich variety of quantum behaviours. %In particular, they exhibit effects linked to quantum interference, tunnelling, and entanglement.
Both discrete-time~\cite{Aharonov93} and continuous-time variants~\cite{Farhi98} can perform universal quantum computation~\cite{Childs,discreteUniv}. Furthermore, they display Anderson-like localization~\cite{Anderson58}, and indeed this fact can be used to implement a quantum algorithm for search on a spatial lattice~\cite{Search}, offering quadratic speedup over the best known classical algorithms.

Quantum walk based schemes for the generation and distribution of entanglement are not new~\cite{Entanglement}. However, the current literature mostly focuses on generating entanglement between the walker's external degrees of freedom and the coin driving its dynamics or, in two-dimensional quantum walks, the two motional states of a given walker~\cite{Carlo}. 

In this paper we present a conceptually new way of using quantum walks to generate entangled states, which we dub the \emph{interacting quantum walk}. In this paradigm, the walker interacts with a collection of qubits, located at the sites of a general lattice, by means of a controlled-Z interaction. Using an extensive computer simulation of the dynamics of both the lattice of qubits and the walker, we show that this scheme is effective in generating multipartite entanglement between the qubits in the walker itself. Moreover, interesting and useful forms of entangled states involving the lattice qubits alone can be produced by arranging suitable interactions with the hovering walker.

Experimentally, considerable success has been found in linear-optics implementations of quantum random walks based on integrated photonics circuits. Recent advances in this context have allowed for the management of disparate degrees of freedom of photonic information carriers, which would allow for the  implementation of controlled interactions at the basis of our proposal. Our scheme, thus, holds the potential to drive the experimental research in quantum random walks well into the domain of entanglement generation and distribution. 

The remainder of this paper is organised as follows: in Sec.~\ref{sec:background} we introduce the formalism for the description of quantum walks in discrete time, and present the protocol for multipartite entanglement distribution that we have designed. Sec.~\ref{sec:results} is dedicated to the presentation of the main results of our computer simulations, and the characterisation of the quality of the states produced through the proposed method. Finally, in Sec.~\ref{sec:conclusion} we draw our conclusions and motivate further investigations. 

\section{\label{sec:background}Background}
\subsection{Discrete-time quantum walks}

\begin{figure}[!t]
	\includegraphics[width=0.4\textwidth] {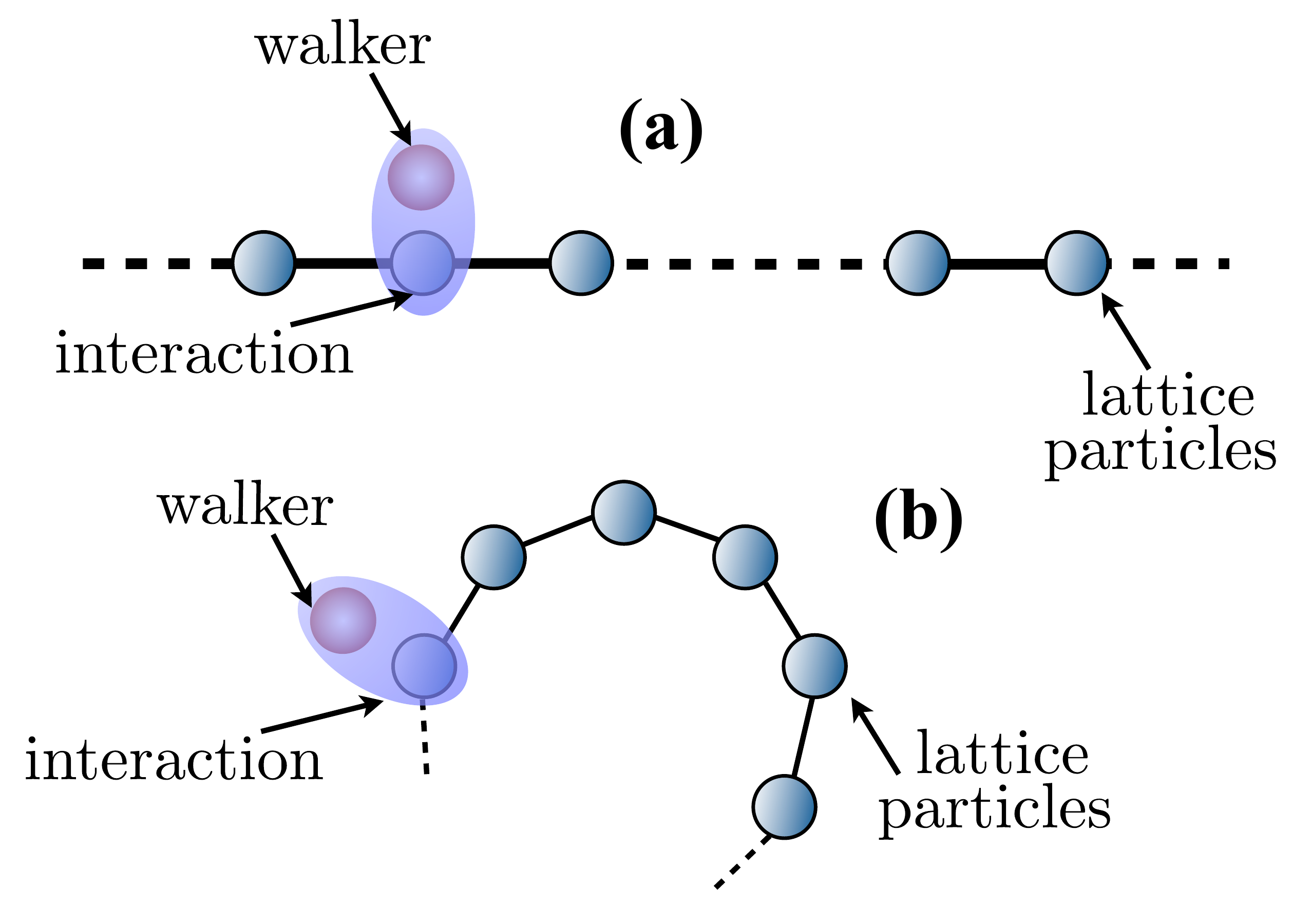}
       \caption{Illustration of the scheme of principle of the situation we consider. A quantum walker moves across the sites of a given lattice according to the `coin tossing' operation defined in the main text. The walker interacts with the `internal' degree of freedom of the particle occupying a given lattice site. By hovering over the lattice, the walker brings about information on the state of the lattice particles, and entangles them in a multipartite fashion. In panel {\bf (a)} [{\bf (b)}] we show the process on a path graph $L_\infty$ [cyclic graphs $C_\infty$].}
        \label{fig:graphs}
\end{figure}

The state space of a quantum walk on a graph $G=(V,E)$ is a bipartite Hilbert space $\mathcal{H}_{W} = \mathcal{H}_{P}\otimes \mathcal{H}_{C}$ consisting of a \emph{position space}
\begin{align}
    \mathcal{H}_{P} := \text{span}\left(\{\ket{v \in V}\}\right)\cong\mathbb{C}^{|V|}.
\end{align}
and a \emph{coin space}
\begin{align}
\mathcal{H}_{C} := \text{span}(\{\ket{d\in [\Delta]}\})\cong\mathbb{C}^{\Delta},
\end{align}
where $\Delta=\max_{v\in V}[\text{deg}(v)]$ is the maximum degree of the graph.
The quantum walk evolves by successive alternating applications of a coin operator $\hat{C}:\mathcal{H}_C\rightarrow \mathcal{H}_C$, followed by a shift operator $\hat{S}:\mathcal{H}_W\rightarrow \mathcal{H}_W$, which moves the walker to another vertex of the graph, dependent on the coin state. 

In this paper we consider quantum walks on path graphs $L_n$ and cycle graphs $C_n$. These topologies utilize the shift operators
\begin{equation}
\begin{aligned}
\hat{S}_{L_n}&:=\ket{n-1}\bra{n-1}_P\otimes \ket{0}\bra{1}_C + \ket{0}\bra{0}_P\otimes \ket{1}\bra{0}_C\\
&+\left(\sum_{i=1}^{n-1}\ket{i-1}\bra{i}_P\right)\otimes \ket{0}\bra{0}_C\\
&+\left(\sum_{i=0}^{n-2}\ket{i+1}\bra{i}_P\right)\otimes \ket{1}\bra{1}_C,
\end{aligned}
\end{equation}
and
\begin{equation}
\begin{aligned}
\hat{S}_{C_n}&:=\sum_{i=0}^{n-1}\Big(\ket{(i-1) \text{ mod } n}\bra{i}_P\otimes \ket{0}\bra{0}_C\\
&+ \ket{(i+1) \text{ mod } n}\bra{i}_P\otimes \ket{1}\bra{1}_C\Big),
\end{aligned}
\end{equation}
respectively. Note the periodic boundary condition implemented by `coin flip' added to the path graph shift operator to maintain unitarity.

\subsection{Interacting quantum walks}

We now pass to the analysis of the interacting quantum walks at the core of our investigation. Given a quantum walk on a simple graph $G$, we associate a qubit with each one of its $n$ vertices. Then, at each step of the walk, a sequence of controlled-Z operators are performed, each with the coin as control and the qubit at one of the vertices as target. 

Schemes where the walk evolution is history dependent are explored in \cite{memorywalks}. These are similar to our interacting walk scheme only in that they associate qubits with vertices, but this is so as to record where the walker has been so far on the graph.

\begin{figure*}[t]
	\includegraphics[width=2\columnwidth]{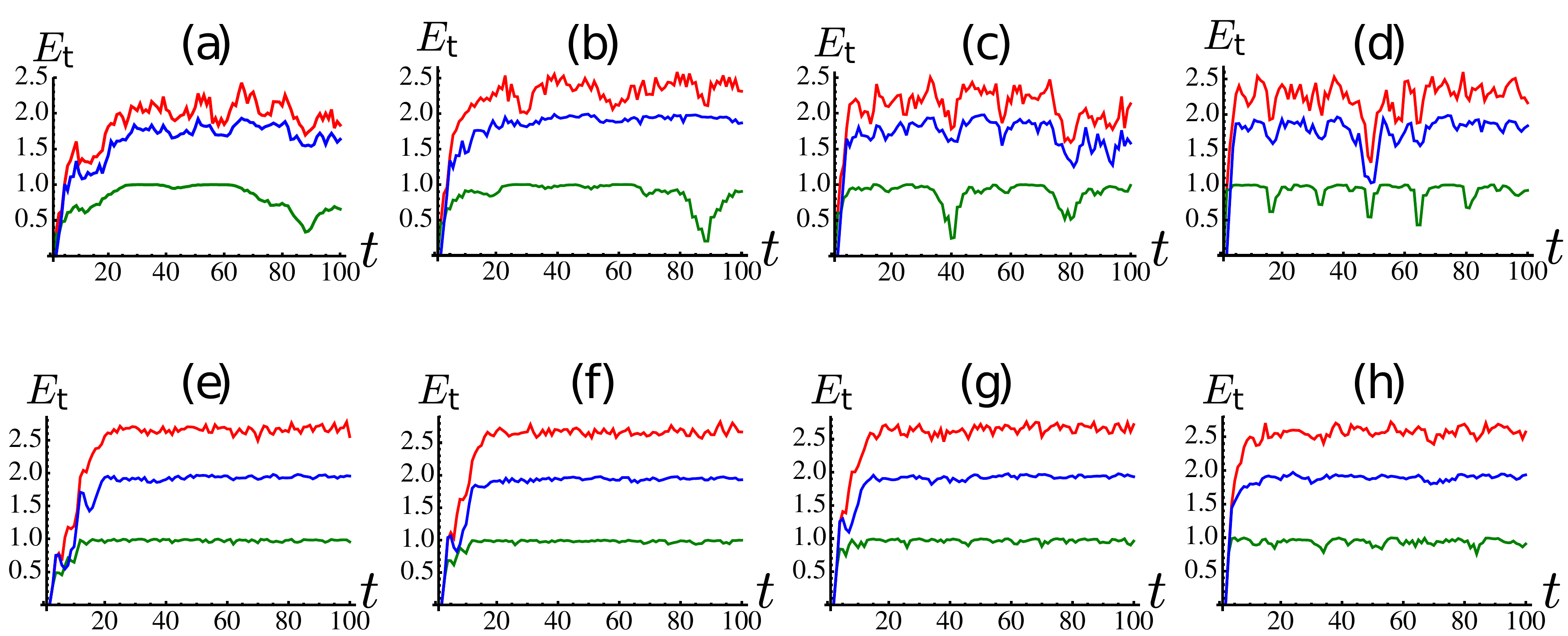}
       \caption{Von Neumann entropy of the reduced state of the vertex qubits, the coin qubit, and the walker system (topmost to lowermost curve respectively). We plot $E_t$ against the discrete evolution time for a cyclic lattice of four sites [{\bf (a)}-{\bf (d)}] and an equal-size linear one [{\bf (e)}-{\bf (h)}]. We have used the coin parameters $(\theta,\phi_1,\phi_2)\in\{(3\pi/20,0,7\pi/20),
       	(\pi/5,0,\pi/5),
       	(\pi/4,0, 2\pi/5),
       	(2\pi/5, 0, 3\pi/10)\}$. As the time $t$ only takes discrete values, the continuous lines are drawn as guidance to the eye.}
        \label{fig:EntropyInsieme}
\end{figure*}

Formally, the total system's Hilbert space is
\begin{align}
\mathcal{H}_{\text{IW}}&=\mathcal{H}_{\text{W}}\otimes(\mathbb{C}^2)^{\otimes n},
\end{align}
and each step of the evolution is augmented with the application of the $\hat{\mathcal{Z}}$ operator, defined as
\begin{align}
\hat{\cal Z}_{i}=\ket{i}\bra{i}_P\otimes\hat Z_{C,i},
\end{align}
where $\hat Z_{C,i}=\left[\ket{0}\bra{0}_C\otimes\mathbbm{1}_i+\ket{1}\bra{1}_C\otimes\hat\sigma_{z,i}\right]$ is a controlled-Z operation with the coin as a control and the $i^{\rm th}$ particle of a given lattice as the target, $\sigma_{z,i}$ is the $z$-Pauli operator for the qubit at vertex $i$ of the graph, and $\mathbbm{1}_i$ is its identity operator. The evolution operator of the interacting walk is then
\begin{equation}
\hat{U}=\left[\prod_{i=0}^{n-1} \hat{\cal Z}_i\right]  (\hat{S}_G \otimes \mathbbm{1}_G)  (\mathbbm{1}_{P}\otimes\hat{C}\otimes \mathbbm{1}_G),
\end{equation}
for $\mathbbm{1}_{P}, \mathbbm{1}_G$ the identity operators on walker position and vertex qubit spaces respectively. In other words, at each step of the walk, the controlled-Z operator is applied with the coin qubit as control, and with the vertex qubit on the walker's current position as the target.
We study the dynamics resulting from the application of this propagator with the aim of generating entangled states of the qubits at the vertices of a given graph. To this end, we consider the reduced state of the vertex qubit system $\rho_{G}(t)=\text{Tr}_{P,C}(\ket{\psi(t)}\bra{\psi(t)}_{\rm IW})$ with  $\ket{\psi(t)}_{\rm IW}=\hat{U}^t\ket{\psi(0)}_{\rm IW}$ the state of the overall system after the walker has made $t$ steps, and investigate the (structure of) entanglement shared by the qubits at the vertices of the graph at hand. Clearly enough, ${\rm Tr}_{P,C}(\cdot)$ stands for the partial trace over the walker coin and position degrees of freedom. 

It can be seen that we have an effective {\it control knob} over the evolution of the walker in the form of the coin operator $\hat{C}$, which, as previously stated, can be any operator in $\text{SU}(2)$. Up to an irrelevant global phase, the most general form of coin operator reads
\begin{equation}
\hat{C}(\theta,\phi_1,\phi_2)=
\begin{pmatrix}
e^{-\frac{i}{2}(\phi_1+\phi_2)}\cos\frac{\theta}{2}&-e^{\frac{i}{2}(\phi_2-\phi_1)}\sin\frac{\theta}{2}\\
e^{\frac{i}{2}(\phi_1-\phi_2)}\sin \frac{\theta}{2}&e^{\frac{i}{2}(\phi_1+\phi_2)}\cos \frac{\theta}{2}
\end{pmatrix},
\end{equation}
which leads to the parameterisation $\hat U\to\hat U(\theta,\phi_1,\phi_2)=\left[\prod_{i=0}^{n-1} \hat{\cal Z}_i\right]  (\hat{S}_G \otimes \mathbbm{1}_G)  (\mathbbm{1}_{P}\otimes\hat{C}(\theta,\phi_1,\phi_2)\otimes \mathbbm{1}_G)$. 
Needless to say, the effect of the evolution induced by the walk operator $\hat U(\theta,\phi_1,\phi_2)$ depends on the initial state of the vertex qubit system. The particular choice of coin-particle unitary evolution suggests the preparation $\ket{+}^{\otimes|V(G)|}$ for the vertex qubits, where $\ket{+}=(\ket{0}+\ket{1})/\sqrt 2$ is the $+1$ eigenstate of $\hat\sigma_x$. In our analysis, we consider the walk initial state $\ket{0}\ket{0}\in \mathcal{H}_P \otimes \mathcal{H}_C$, where in the path graph, position $0$ corresponds to the leftmost vertex.
\section{\label{sec:results}Analysis of dynamics and entanglement sharing structure}
\subsection{Entanglement across bipartitions of the system}
\label{subsection:entanglementacrossbipartitionsofthesystem}

We have investigated entanglement in the various bipartitions of the system. The state of the overall system is pure, so we will use the von Neumann entropy to quantify entanglement across a given bipartition. This is defined as 
\begin{equation}
E_{t}=-{\rm Tr}[\rho_s\log_2\rho_s],
\end{equation}
with $\rho_s$ the reduced density matrix of one of the subsystems $s=P,C,G$ of a given bipartition. In this way we have considered the entanglement between the walker and the subsystem consisting of the coin and the vertex qubits. In this case then $\rho_s={\rm Tr}_{C,G}[\ket{\psi(t)}\bra{\psi(t)}_{\rm IW}]$ with $G=L_n$ ($G=C_n)$ for a linear (cyclic) graph. The results are shown in Fig.~\ref{fig:EntropyInsieme} (blue (middle) curves in each panel) for lattices of both configurations, $n=4$ sites, and four different sets of coin-operation parameters. Quantitatively, the amount of entanglement is similar for the two configurations, and rises almost to its maximum value of 2 ebits after a short time (the linear configuration requiring a relatively longer time), regardless of the coin-operation parameters. However, a linear lattice sets a steadier degree of entanglement in the time-asymptotic limit: the blue (middle) curves in Fig.~\ref{fig:EntropyInsieme} {\bf (e)}-{\bf (h)} display negligible fluctuations around the asymptotic value. 

A similar behavior is observed for the amount of entanglement between the coin's degree of freedom and the subsystem consisting of the walker and vertex qubits, i.e. for $\rho_s={\rm Tr}_{P,G}[\ket{\psi(t)}\bra{\psi(t)}_{\rm IW}]$ [cf. green (lowermost) curves in Fig.~\ref{fig:EntropyInsieme}]. Obviously, in this case the maximum degree of entanglement that can be achieved is 1 ebit, as the coin is a qubit. However, the temporal behavior of entanglement is much more unsteady than that between the walker and the rest of the system.  

Finally, when focusing on the entanglement between the vertex qubits and the joint system composed of the walker and coin, we find that the cyclic lattice configuration resembles closely the entanglement between the coin and the rest of the system. As the dimension of the vertex qubit system is $2^4=16$, the maximum amount of entanglement that can be shared across such a bipartition is 4 ebits, which is never achieved for the coin parameters used in our simulations. Our analysis demonstrates the tripartite nature of the walker-coin-lattice state (owing to the inseparable nature of the state of every bipartition being considered out of such a grouping), as well as suggesting the negligibility of the entanglement between the walker and coin degrees of freedom in the cyclic configuration.  

\begin{figure}[t]
	\includegraphics[width=0.9\columnwidth]{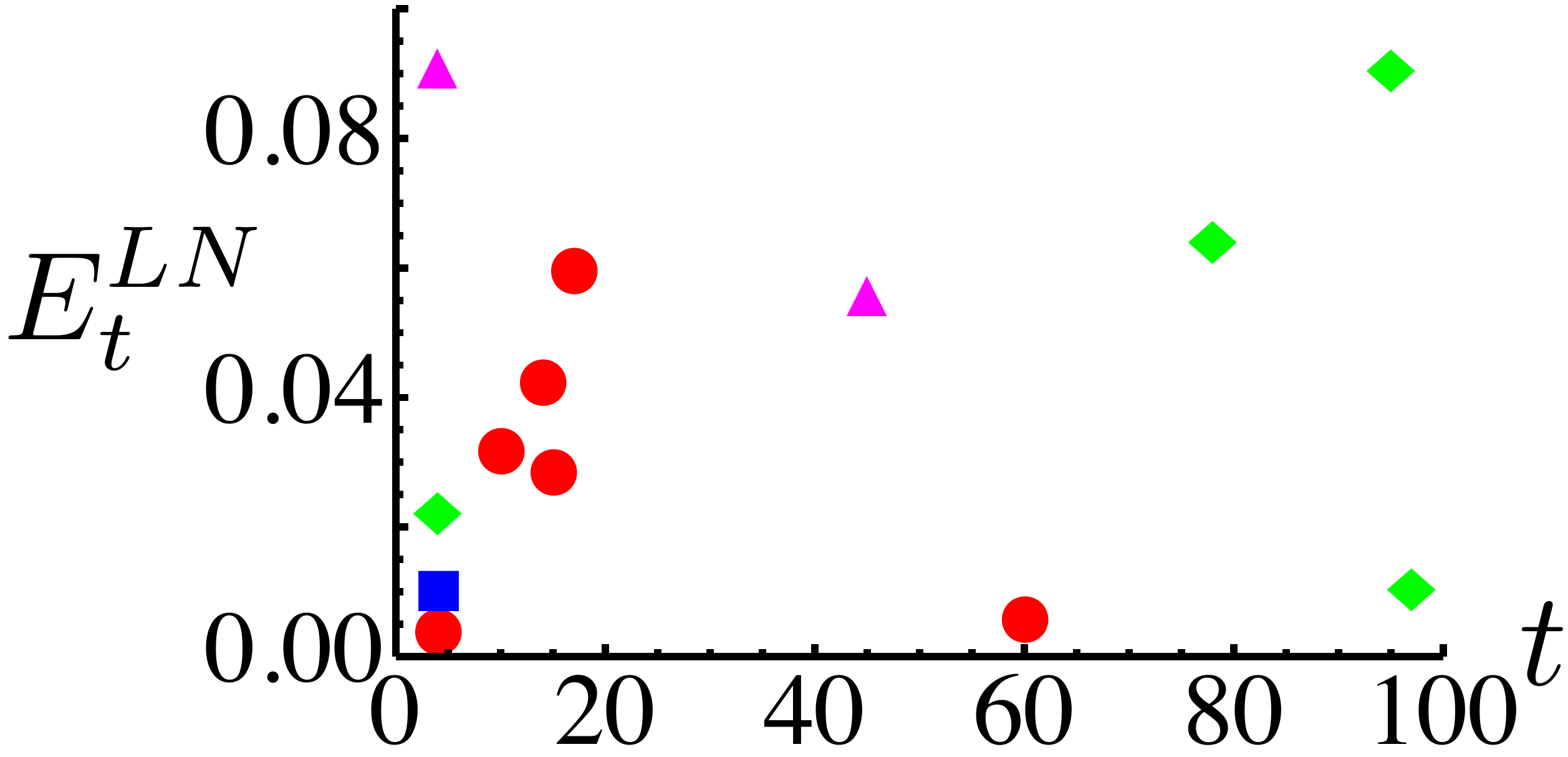}
       \caption{Logarithmic negativity of the walker-coin state $\rho_{PC}$ plotted against the discrete interaction time $t$ for the four sets of coin parameters considered in Fig.~\ref{fig:EntropyInsieme} (plotted in the order listed in Fig.~\ref{fig:EntropyInsieme} with circles, squares, diamonds, and triangles respectively) and a cyclic lattice of $n=4$ sites. We plot only the non-zero values of $E^{LN}_t$.}
        \label{lognegWCCyclic}
\end{figure}

\begin{figure}[t]
	\includegraphics[width=\columnwidth]{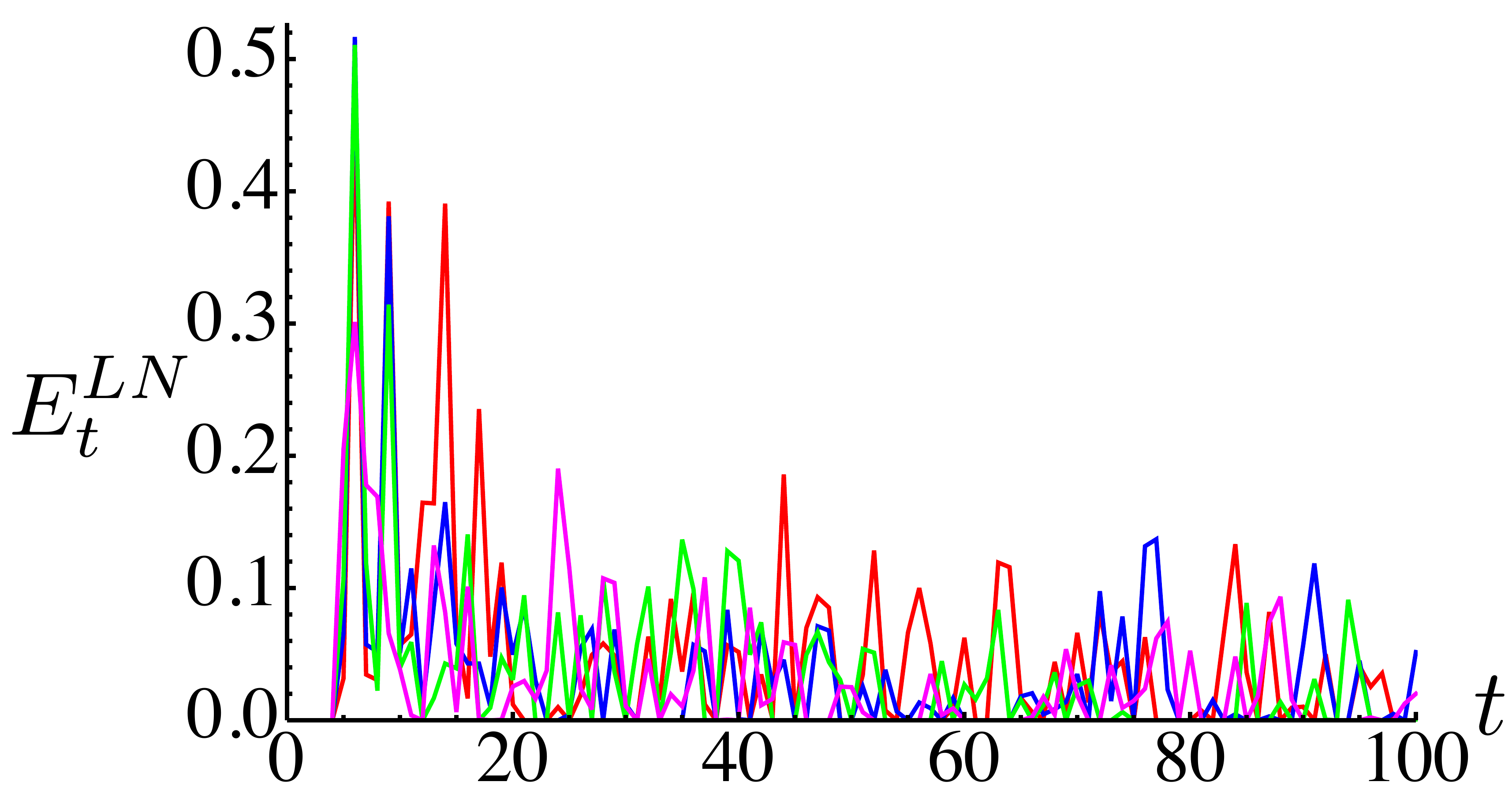}
       \caption{Logarithmic negativity of the walker-coin state $\rho_{PC}$ plotted against the discrete interaction time $t$ for the four sets of coin parameters considered in Figs.~\ref{fig:EntropyInsieme} and a linear lattice of $n=4$ sites. We plot only the non-zero values of $E^{LN}_t$. As the evolution time is discrete the continuous lines should be interpreted only as guidances to the eye.}
        \label{lognegWCLinear}
\end{figure}

When assessing the entanglement brought about by reductions of the overall system, we find that the walker is only very weakly entangled with the coin, in the case of a cyclic lattice~\cite{noteBound}. On the other hand, a more substantial degree of entanglement can be found in the linear configuration. This can be seen by considering the reduced state $\rho_{PC}={\rm Tr}_{C_n}[\ket{\psi(t)}\bra{\psi(t)}_{\rm IW}]$ and evaluating its degree of entanglement  using the logarithmic negativity~\cite{vidal}
\begin{equation}
E^{LN}_t=\max[0,\log_2\|\rho^{PT}_{PC}\|_1],
\end{equation}
which is based on the negativity of partial transposition (NPT) criterion~\cite{npt,Horodecki}. Here $\rho^{PT}_{PC}$ is the partial transposition of $\rho_{PC}$ with respect to the coin subsystem, and $\|X\|_1=\text{Tr}\left[\sqrt{X^\dagger X}\right]$ is the Schatten 1-norm defined on a linear operator $X$. The logarithmic negativity was proven to be a proper entanglement monotone in Ref.~\cite{Plenio}. The results of such calculations are presented in Figs.~\ref{lognegWCCyclic} and~\ref{lognegWCLinear}, which show $E^{LN}_t$ for the cyclic and linear lattice respectively, and all the coin-operation sets of parameters considered in our simulations [cf. Figs.~\ref{fig:EntropyInsieme}]. 

\begin{figure}[b!]
	\includegraphics[width=\columnwidth]{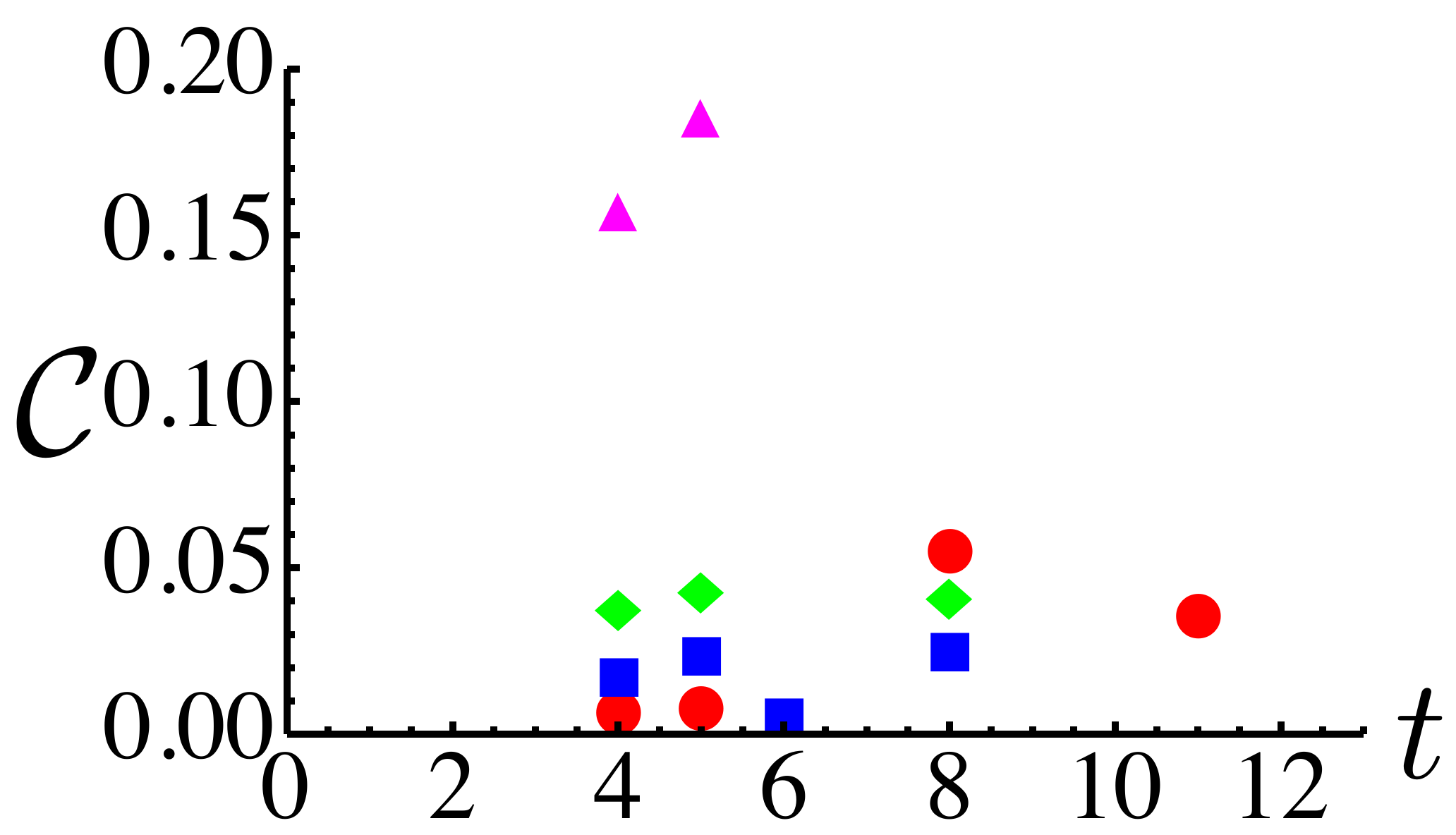}
       \caption{Four-partite concurrence of the state of the vertex qubits for the four sets of coin-operation parameters used in our simulations.}
        \label{fig:NoPostselection}
\end{figure}

\begin{figure*}[t]
	\includegraphics[width=1.9\columnwidth]{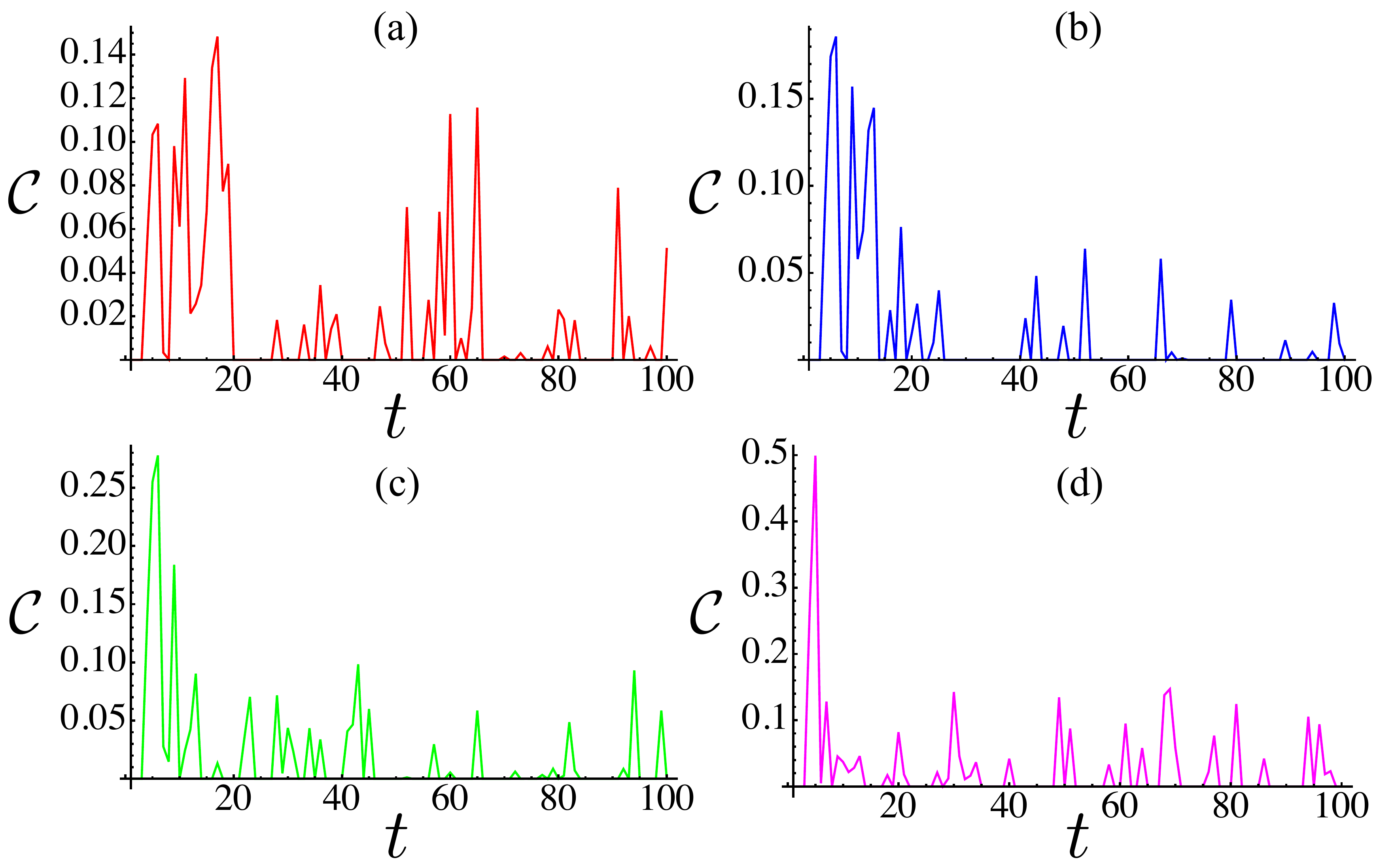}
       \caption{Four-partite concurrence of the conditional state of the vertex qubits obtained by discarding the state of the walker and projecting the coin onto the state corresponding to $\nu=0$ and $\mu=0$. We have considered the four sets of coin-operation parameters used in our simulations.}
        \label{fig:Postselection}
\end{figure*}

\begin{figure*}[!t]
	\includegraphics[width=1.9\columnwidth]{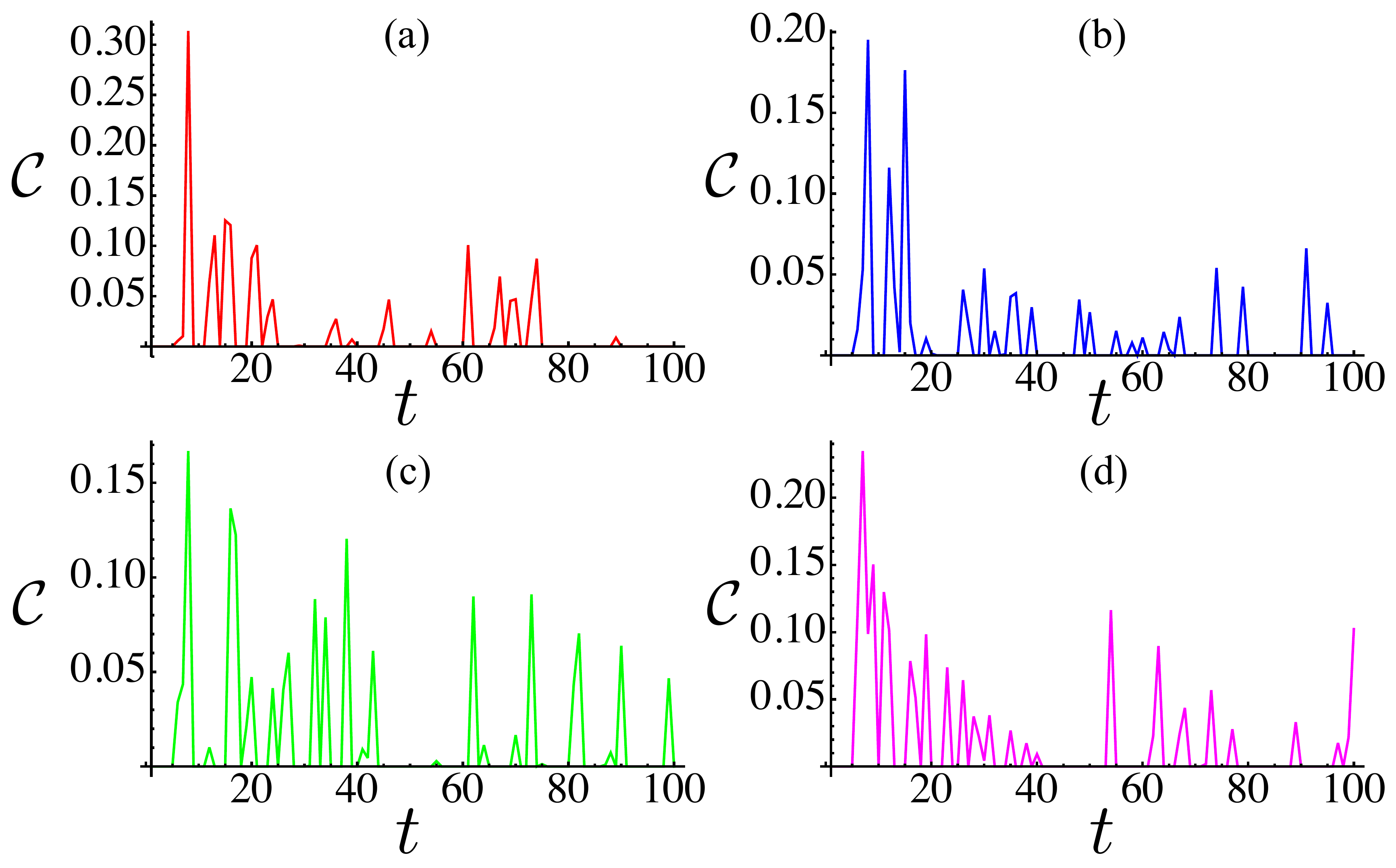}
	\caption{Four-partite concurrence of the conditional state of the vertex qubits obtained by discarding the state of the walker and projecting the coin onto the state corresponding to $\nu=0$ and $\mu=\pi/2$. We have considered the four sets of coin-operation parameters used in our simulations.}
	\label{fig:PostselectionPiOver2}
\end{figure*}

\begin{figure*}[!t]
	\includegraphics[width=1.9\columnwidth]{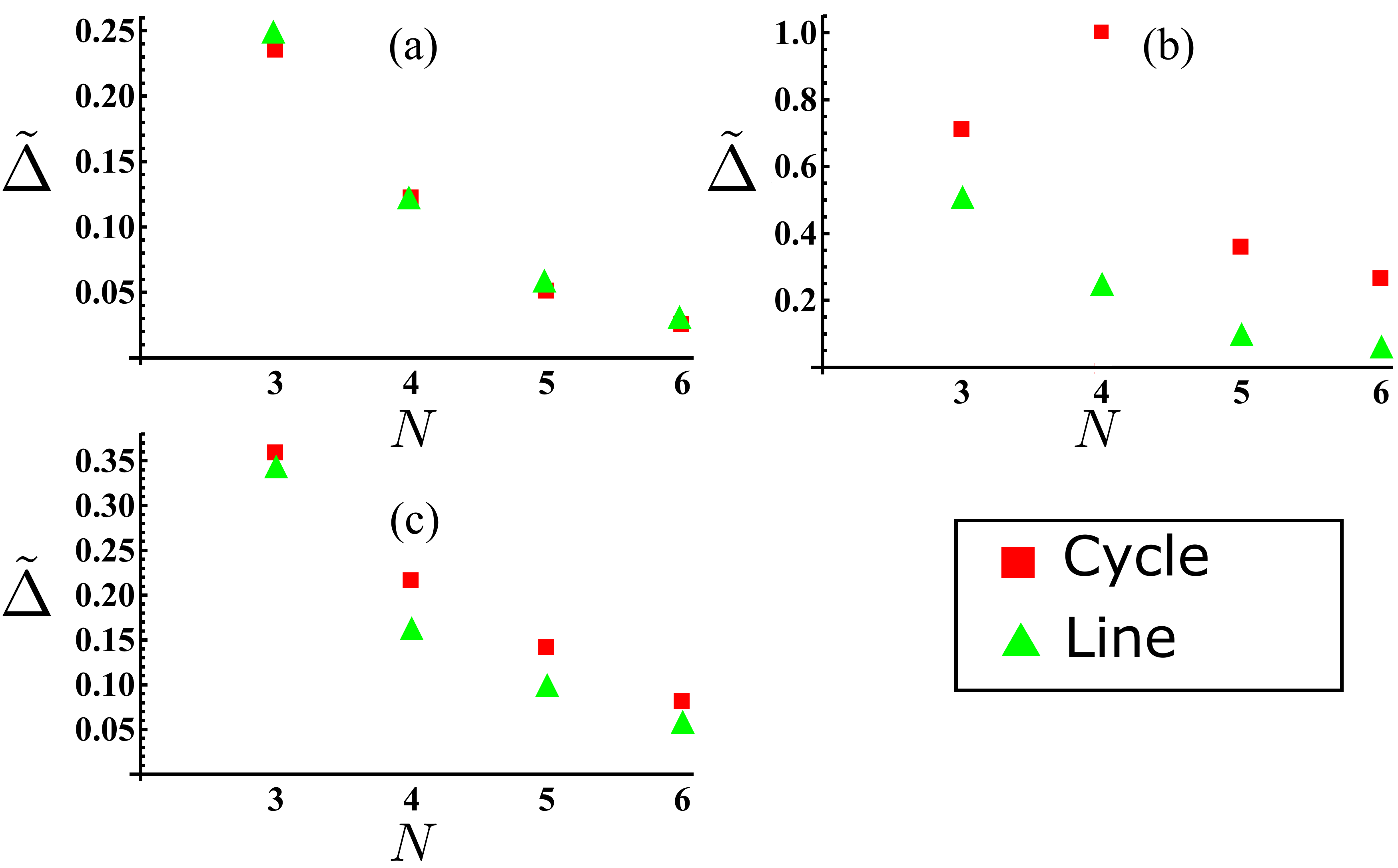}
	\caption{Comparison of minimum achieved trace distance from states of interest, $\widetilde{\Delta}$ for \textbf{(a)} $\rho_E=\ket{\text{GHZ}_N}\bra{\text{GHZ}_N}$, \textbf{(b)} $\rho_E=\ket{C_N}\bra{C_N},\ket{L_N}\bra{L_N}$, and \textbf{(c)} $\ket{\text{W}_N}\bra{\text{W}_N}$.}
	\label{fig:comparetds}
\end{figure*}%

\begin{figure*}[t!]
	\includegraphics[width=1.9\columnwidth]{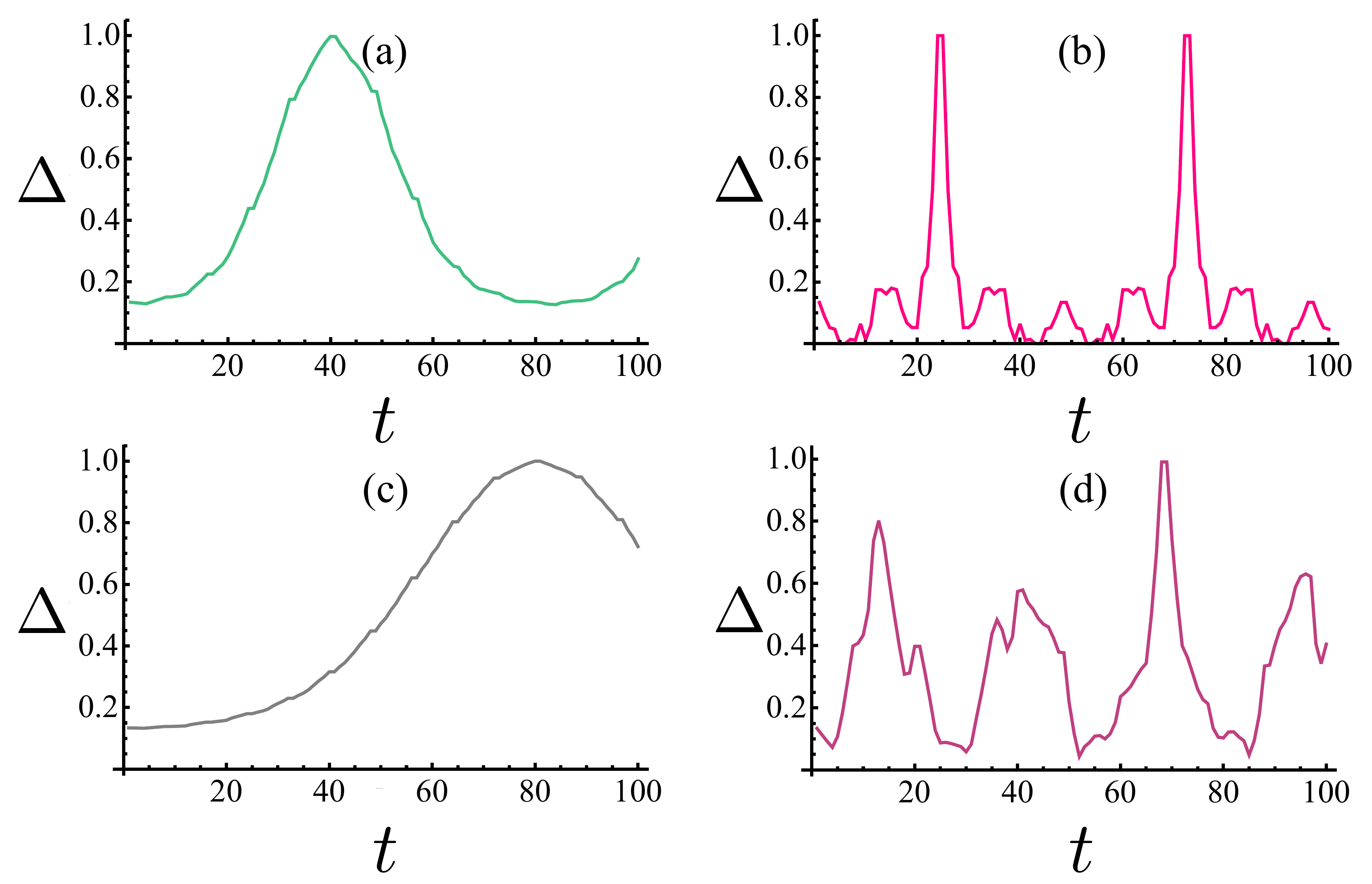}
	\caption{Inverse trace distance $\Delta(\rho_G(t), \ket{C_4}\bra{C_4})$ for four best performing parameter sets $(\theta,\phi_1,\phi_2)$: \textbf{(a)} $(\pi/10,0,0)$ \textbf{(b)} $(\pi/2,0,\pi/2)$, \textbf{(c)} $(\pi/20,0,0)$, \textbf{(d)} $(7\pi/20,0,0)$.}
	\label{fig:graphstatetd}
\end{figure*}%

\subsection{Entanglement between vertex qubits}
\label{subsection:entanglementbetweenlatticequbits}
When looking into the entanglement actually shared by the vertex qubits, the problem of choosing a suitable quantifier of multipartite entanglement arises. While no unique indicator is currently at hand, useful characterisations can be made using situation-specific figures of merit. One is provided by the $n$-partite concurrence, which is an extension of the widely used bipartite concurrence~\cite{Wootters}. For an $n$-qubit density matrix $\rho$ this is defined as
\begin{align}
\mathcal{C}(\rho)=\max\left[0, \sqrt{\lambda_1}-\sum_{j=2}^{2^n}\sqrt{\lambda_j}\right],
\end{align}
where $\lambda_1 \ge \lambda_2 \ge \dots \ge \lambda_{2^n}$ are the eigenvalues of the operator
\begin{align}
\tilde\rho=\rho\hat S_y\rho^*\hat S_y.
\end{align}
with $\hat S_y=\bigotimes_{j=1}^N\hat{\sigma}_y$, $\rho^*$ the conjugate of $\rho$, and $\hat\sigma_y$ the $y$-Pauli matrix. This measure is appealing due to its computationally handy form and its sensitivity to global entanglement. In fact, it is exactly zero if any qubit is separable from the rest of a system. Its non-nullity, though, is only a sufficient condition for multipartite entanglement as there are multipartite entangled states with vanishing $n$-concurrence, such as the $n$-qubit W-states \cite{Horodecki}. 
With such proviso, the outcomes of the evaluation of ${\cal C}(\rho_{G})$ are shown in Fig.~\ref{fig:NoPostselection}, where only the 
values associated with the linear lattice configuration are shown, those of the cyclic one being identically null. The four-partite entanglement within the lattice manifests only at specific instants of the evolution, and remains at low values for most of the coin-parameter set. 

The situation can be changed rather significantly by exploiting the entanglement between the walker-coin subsystem and the vertex qubits.

 By post-selecting the state of the vertex qubits upon projection of the state of the other systems, we have, in principle, the possibility to enhance the degree of multipartite entanglement shared by the particles at the vertices of the lattice being considered. In order to provide evidence of such a possibility, here we address the case corresponding to tracing out the state of the walker and projecting the coin onto the state
\begin{equation}
\ket{\Sigma}_C=\cos\mu\ket{0}_C+e^{-i\nu}\sin\mu\ket{1}_C.
\end{equation}
An extensive numerical exploration of the behavior of the four-partite concurrence performed by taking $\mu\in[0,\pi]$ and $\nu\in[0,\pi/2]$, and using the four coin-parameter sets used in the simulations studied herein shows that, while such a procedure is ineffective for a cyclic lattice, it works well for a linear one. The maximum of ${\cal C}(\rho_{L_4})$ is found for projections of the coin onto its computational states, i.e. for $\nu=0$ with $\mu=0$ or $\pi/2$. The temporal behavior of the four-partite concurrence for such projections performed on the 
linear lattice configuration is shown in Figs.~\ref{fig:Postselection} and~\ref{fig:PostselectionPiOver2}, which demonstrate the improvement achieved upon implementation of the conditional strategy, and the qualitatively similar behavior observed for the two projections being considered. 

The remainder of our analysis aims at shedding light on the actual form of the entangled states that we achieve. 

\subsection{\label{subsection:tracedistancefromentangledstates}Trace distance from entangled states}
In order to gather information on the form of the state produced by the interacting quantum walk protocol introduced here, we quantify the closeness of $\rho_{G}$ to a number of ``archetypal'' multipartite entangled states. We use the quantity $\Delta(\rho,\sigma)=1-\delta(\rho,\sigma)$, which is defined in terms of the \emph{trace distance} between two generic density matrices $\rho,\sigma$, 
\begin{equation}
\delta(\rho,\sigma)=\frac{1}{2}\|\rho-\sigma\|_1= \frac{1}{2}\sum_{i}|\epsilon_i|^2,
\end{equation}
where $\epsilon_i$ are the eigenvalues of the operator $\rho-\sigma$.  
We consider GHZ and W states~\cite{Horodecki}, as well as graph states corresponding to the lattice arrangements at hand in our analysis. Graph states are generalisations of cluster states, which are the key ingredient in measurement based quantum computation~\cite{Raussendorf}. Given a graph $G=(V,E)$, the corresponding graph state is defined as
\begin{align}
\ket{G}=\prod_{\{i,j\}\in E}\hat Z_{i,j}\ket{+}^{\otimes |V|},
\end{align}
where $\hat Z_{i,j}$ is the controlled-Z operator with qubit $i$ as control and qubit $j$ as target, and $\ket{+}=({\ket{0}+\ket{1}})/{\sqrt{2}}$.

We begin our characterisation of the qubit state resulting from the interacting walk by calculating its trace distance from GHZ, W and corresponding graph states $\ket{C_n},\ket{L_n}$. 
We see from Fig.~\ref{fig:comparetds}, where we plot the quantity 
  $\widetilde{\Delta} = \max_{\theta,\phi_1,\phi_2,t}\Delta\left(\rho_{G}(t),\rho_E\right)$
(with $\rho_E=\ket{\text{GHZ}_n}\bra{\text{GHZ}_n},\ket{\text{W}_n}\bra{\text{W}_n}$, and $\ket{G}\bra{G}$ for
 $G=C_n,L_n$), that the quantum walk on the four-vertex cycle graph generates a perfect $\ket{C_4}\bra{C_4}$ state at {\it some} instant of time.
To investigate this further, in Fig.~\ref{fig:graphstatetd} we plot the evolution of $\Delta(\rho_{C_4},\ket{C_4}\bra{C_4})$ for the sets of coin parameters corresponding to the highest values achieved during the first $100$ time steps. 
The highest value attained is exactly $1.0$, and is found at $t=24$ using $\hat{\text{C}}(\frac{\pi}{2},0,\frac{\pi}{2})$. We also appreciate that at no point in the first $100$ steps is the state of the vertex qubit system close to a W or GHZ state, regardless of the coin operator and graph topology. From this it is also clear that the linear lattice configuration performs markedly worse than the cyclic one.

\section{\label{sec:conclusion}Conclusion}

We have investigated a scheme for generating multipartite entanglement which we call the \emph{interacting quantum walk}. As described above, this is where a quantum walker evolves on a graph and interacts with qubits located on the vertices via a phase shift interaction. We have performed a computer simulation of the walker and vertex qubit system dynamics, and have shown that the scheme can be used to generate multipartite entanglement between the qubits, as well as between the qubits and the walker and coin. In particular, we have shown that it can be used to prepare $4$ qubit cyclic cluster states, of use in measurement based quantum computation.

While here we have only considered linear and cyclic graph structures, a natural extension would be to consider different graph topologies and the effect they have on the amount of entanglement generated by the system. It is notable that the states generated by the interactions are never close to GHZ or W states. It would be interesting to find which graphs cause these states to be generated.

A pre-requisite for performing measurement based quantum computing is a reliable method for generating large graph states. In~\cite{Fusion}, Browne and Rudolph discuss ``fusion'' operators for chaining together small graph states to generate larger ones: an interesting direction of research would be to adapt these methods to situations where different clusters, each generated through the protocol presented here, are joined together via effective fusion gates.

Finally, an exciting future direction would be to implement the interacting quantum walk in a laboratory setting. Experimental implementations of quantum walks are becoming more mature, and it could potentially be possible to engineer the controlled phase interaction required for our protocol. In particular, the use of superlattices in ultra-cold atomic settings, a scenario where quantum walks have been controllably implemented, could be a feasible avenue to explore~\cite{mechede}. This is particularly interesting in light of the fact that schemes for the generation of cluster states in superlattices arrangements have been put forward already in the past~\cite{mauro}. 

\acknowledgments
JL thanks Dan Browne for useful discussions, and acknowledges the EPSRC Centre for Doctoral Training in Delivering Quantum Technologies at UCL for financial support. MP acknowledges the John Templeton Foundation (grant ID 43467),  the UK EPSRC (EP/M003019/1), and the EU FP7 grant TherMiQ (Grant Agreement 618074) for financial support.

\end{document}